\documentclass[12pt]{article}
\pdfoutput=1
\usepackage[reqno]{amsmath}
\usepackage{mathrsfs}
\usepackage{amssymb}

\usepackage{rotating} %for landscape format

\usepackage{array}
\usepackage{float}
\usepackage{color}
\usepackage{graphicx}% Include figure files
\usepackage{a4}
\usepackage{a4wide}
\usepackage{wasysym}
\usepackage{cite}
\def\gs{\mathrel{
   \rlap{\raise 0.511ex \hbox{$>$}}{\lower 0.511ex \hbox{$\sim$}}}}
\def\ls{\mathrel{
   \rlap{\raise 0.511ex \hbox{$<$}}{\lower 0.511ex \hbox{$\sim$}}}}

\newcommand{\be}{\begin{eqnarray}}
\newcommand{\ee}{\end{eqnarray}}
\newcommand{\beq}{\begin{equation}}
\newcommand{\eeq}{\end{equation}}

\newcommand{\beqn}{\begin{eqnarray}}
\newcommand{\eeqn}{\end{eqnarray}}

\newcommand{\ka}{\kappa}

\newcommand{\eps}{\mbox{$\epsilon$}}

\begin{document}

%For feynmf-Package:
\setlength{\unitlength}{1mm}

\begin{titlepage}
\title{\vspace*{-3.0cm}
\hfill 
{\small 
\begin{tabular}{l}
SI-HEP-2014-08\\
QFET-2014-04
\end{tabular}}\\[15mm]
\bf\Large
Radiative Inflation and Dark Energy RIDEs Again after BICEP2
\\[5mm]\ }

\author{
Pasquale Di Bari$^a$\thanks{email: \tt P.Di-Bari@soton.ac.uk}~~,~~
Stephen F.\ King$^a$\thanks{email: \tt S.F.King@soton.ac.uk}~~,~~
Christoph Luhn$^b$\thanks{email: \tt Christoph.Luhn@uni-siegen.de}~~,~~\\
Alexander Merle$^a$\thanks{email: \tt A.Merle@soton.ac.uk}~~,~~and~~
Angnis Schmidt-May$^c$\thanks{email: \tt Angnis.Schmidt-May@fysik.su.se}
\\ \\
{\normalsize $^a$ \it Physics and Astronomy, University of Southampton,}\\
{\normalsize \it Southampton, SO17 1BJ, United Kingdom}\\
\\
{\normalsize $^b$ \it Theoretische Physik 1, Naturwissenschaftlich-Technische Fakult\"at,}\\
{\normalsize \it Universit\"at Siegen, Walter-Flex-Stra{\ss}e 3, D-57068 Siegen, Germany}\\
\\
{\normalsize $^c$ \it Department of Physics \& The Oskar Klein Centre,}\\
{\normalsize \it Stockholm University, AlbaNova University Centre, SE-106 91 Stockholm, Sweden}
}
\date{\today}
\maketitle
\thispagestyle{empty}

\begin{abstract}
\noindent
Following the ground-breaking measurement of the tensor-to-scalar ratio $r = 0.20^{+0.07}_{- 0.05}$ by the BICEP2 collaboration, we perform a statistical analysis of a model that combines Radiative Inflation with Dark Energy (RIDE) based on the $M^2 |\Phi|^2 \ln \left( |\Phi|^2/\Lambda^2 \right)$ potential  and compare its predictions to those based on the traditional chaotic inflation $M^2|\Phi|^2$ potential. We find a best-fit value in the RIDE model of $r=0.18$ as compared to $r=0.17$ in the chaotic model, with the spectral index being $n_S=0.96$ in both models.
\end{abstract}

\end{titlepage}

%%%%%%%%%%%%%%%%%%%%%%%%%%%%%%%%%%%%%%%%%%%%%%%%%%%%%%%%%%%%%%%%%%%%%%
\section{\label{sec:Introduction}Introduction}
%%%%%%%%%%%%%%%%%%%%%%%%%%%%%%%%%%%%%%%%%%%%%%%%%%%%%%%%%%%%%%%%%%%%%%

Recently the BICEP2 collaboration~\cite{Ade:2014xna} has reported a measurement of the tensor-to-scalar ratio $r = 0.20^{+0.07}_{- 0.05}$ which provides evidence for large-field (e.g.\ chaotic) inflationary models, as opposed to small-field (e.g.\ hybrid) inflationary models (for a review of inflationary models see e.g.~\cite{Lyth:1998xn}). BICEP2 provides the first direct measurement of the large scale $B$-mode polarisation power spectrum. This receives a contribution only from tensor perturbations and, therefore, the detection of a non-vanishing $B$-mode polarisation is direct evidence for the presence of tensor perturbations. More indirectly, temperature anisotropies can also potentially give evidence of tensor perturbations. In this way, prior to BICEP2, the {\em Planck} collaboration had placed an upper bound $r < 0.11\,(95\%~\rm{C.L.})$ assuming no running of the scalar spectral index. The origin of this disagreement could either be some statistical or systematic effect, or perhaps the first evidence for a running spectral index that would relax the {\em Planck} limit to $r < 0.26\,(95\%~\rm{C.L.})$~\cite{Ade:2013zuv}, which is well compatible with the BICEP2 measurement.

The BICEP2 results~\cite{Ade:2014xna} have, within a short time, triggered a series of papers on model updates in the light of the new measurement. Among the models investigated are several scenarios of chaotic inflation~\cite{Harigaya:2014sua,Harigaya:2014qza,Lee:2014spa}, broken primordial power spectrum models~\cite{Hazra:2014aea},
gravity-related scenarios~\cite{Channuie:2013xoa,Joergensen:2014rya,Pallis:2014dma,Bamba:2014jia,Bezrukov:2014nza,Ferrara:2014ima,Kidani:2014pka,Kallosh:2014qta,Hamaguchi:2014mza}, Higgs-related inflation~\cite{Nakayama:2014koa,Cook:2014dga,Germani:2014hqa,Bezrukov:2014bra,Costa:2014lta,Fairbairn:2014nxa,Hamada:2014iga}, scenarios related to supersymmetry~\cite{Dimopoulos:2014boa,Craig:2014rta,Ellis:2014rxa,Lyth:2014yya}, curvaton model~\cite{Byrnes:2014xua}, or natural inflation~\cite{Freese:2014nla,Czerny:2014qqa}. Furthermore, several general analyses of collections of simple models have been presented~\cite{Kobayashi:2014jga,Okada:2014lxa}, general statements about the properties of the inflationary potential are available~\cite{Choudhury:2014kma,Choudhury:2013iaa,Hotchkiss:2011gz,Gong:2014cqa} as well as studies of the consistency of the {\em Planck} and BICEP2 data sets~\cite{McDonald:2014kia,Zhang:2014dxk}, and a general discussion about what we can learn from the exciting new results~\cite{Dodelson:2014exa,Cheng:2014ota,Cheng:2014bma}.  

None of the above studied models of inflation is able to simultaneously account for both inflation in the early Universe and the preponderance of Dark Energy in the Universe at the current epoch~\cite{Kinney:2009vz,Copeland:2006wr}. Motivated by the desire to account for both phenomena, some time ago we proposed a model of Radiative Inflation and Dark Energy (RIDE)~\cite{DiBari:2010wg}. Earlier attempts~\cite{Rosenfeld:2005mt}, based on the ``schizon model''~\cite{Hill:1988bu,Frieman:1991tu,Frieman:1995pm}, were formulated in the framework of $\varphi^4$ chaotic inflation, which was subsequently essentially excluded by WMAP 5-year data~\cite{Hinshaw:2008kr}. The question we considered was whether the nice feature of such models, namely that they naturally generate a pseudo Nambu-Goldstone boson (PNGB), which receives a potential via gravitational effects~\cite{Kallosh:1995hi} and can then be used as quintessence field, could be implemented within a viable model of inflation. 

In this note, following the BICEP2 measurement, we revisit the RIDE
model~\cite{DiBari:2010wg} based on the idea of a massive complex scalar field
$\Phi$ whose mass squared is driven negative close to the Planck scale $M_P$
by radiative effects, leading to a potential of the form\footnote{This
  potential belongs to a class of scenarios recently studied in a systematic
  way in \cite{Martin:2013tda}.} $M^2 |\Phi|^2 \ln \left(|\Phi|^2/\Lambda^2 \right)$
which may be compared to the traditional chaotic $M^2|\Phi|^2$
potential.\footnote{We emphasise that both RIDE and chaotic $\varphi^2$ inflation 
  share the need to forbid a possible quartic term~$(\Phi^\dagger \Phi)^2$, or
  at least suppress it sufficiently.  As this cannot be achieved at the level
  of an effective theory, it is necessary to resort to a concrete model
  realisation. This fact has already been commented in~\cite{DiBari:2010wg},
  and an example of such a framework was also presented where the absence of
  the quartic term was achieved in a supersymmetric context where no $D$-terms
  arise. We refer the inclined reader to~\cite{DiBari:2010wg} for  a more
  detailed discussion.} 
The potential is invariant under a global $U(1)$ symmetry, and the absolute
value of the complex field $\Phi$ plays the role of the inflaton field which
rolls slowly down a simple potential that resembles chaotic inflation for high
field values. Similar to chaotic inflation, the RIDE model leads to
predictions for inflation which were fully consistent with WMAP 7-year
data~\cite{Komatsu:2010fb}, but potentially threatened later on by {\em
  Planck}~\cite{Ade:2013zuv}. However, unlike the chaotic model, at the end of
inflation the inflation field settles at a non-trivial minimum, thereby
breaking the global $U(1)$ and generating a PNGB which receives a small mass
via gravitational effects. The resulting effective potential for the PNGB is
of a form which is suitable for a quintessence field. Thus the PNGB can
explain the existence of Dark Energy. Here we perform a statistical analysis
of the RIDE model and compare its predictions to those of the chaotic
inflation model. We find a best-fit value in the RIDE model of $r=0.18$ as
compared to $r=0.17$ in the chaotic model, with the spectral index being
$n_S=0.96$ in both models. 

The layout of the remainder of this paper is as follows. After reviewing the RIDE model in Sec.~\ref{sec:Model}, we perform a statistical analysis of inflation within the RIDE model in Sec.~\ref{sec:Inflation} and compare its predictions to those of chaotic inflation. We conclude in Sec.~\ref{sec:Conclusions}.

%%%%%%%%%%%%%%%%%%%%%%%%%%%%%%%%%%%%%%%%%%%%%%%%%%%%%%%%%%%%%%%%%%%%%%
\section{\label{sec:Model}Review of the RIDE model}
%%%%%%%%%%%%%%%%%%%%%%%%%%%%%%%%%%%%%%%%%%%%%%%%%%%%%%%%%%%%%%%%%%%%%%

\begin{figure}[t]
\centering
\includegraphics[width=0.48\textwidth]{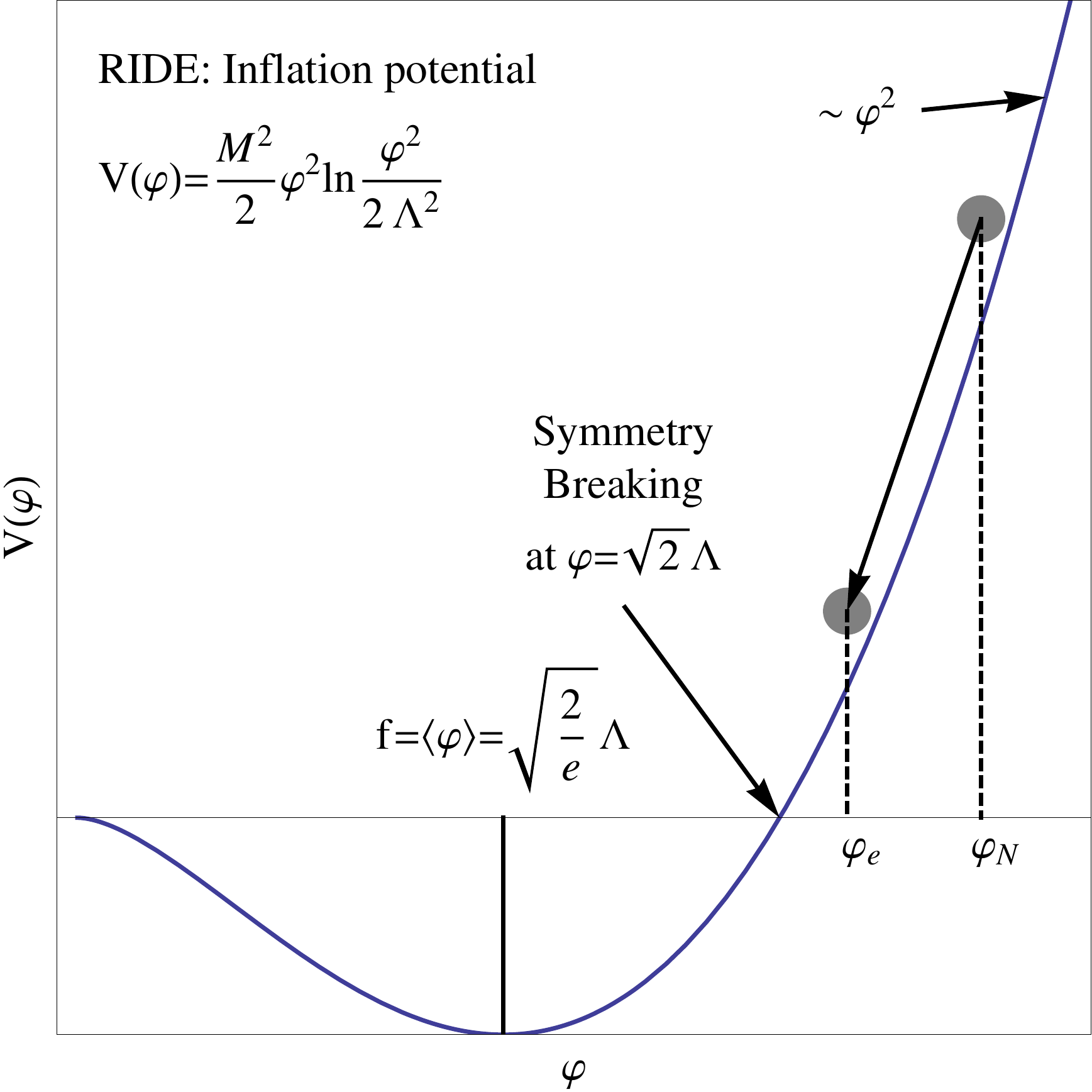}
\vspace*{-4mm}
    \caption{\label{fig:potentials}The shape of the RIDE potential Eq.~\eqref{eq:inf_pot} (with $C=0$), for the inflaton field $\varphi$. Adapted from Fig.~1a of~\cite{DiBari:2010wg}. }
\vspace*{-2mm}
\end{figure}

The model is based on a complex scalar field 
\be
\Phi=\frac{1}{\sqrt{2}} \, \varphi\, e^{i\phi/f} \ ,
\ee
with $f=\langle \varphi \rangle$, whose tree-level potential has a simple quadratic form, $V_0\approx M^2 \Phi^\dagger \Phi$. Including radiative corrections, the mass squared can be driven negative at some scale $\Lambda$ not too far below the Planck scale. The mechanism of radiative symmetry breaking is well-known in the minimal supersymmetric standard model~\cite{Ibanez:1982fr}, where the Higgs mass squared is driven negative at the TeV scale. Similarly, radiative symmetry breaking can play an important role in different contexts~\cite{Varzielas:2006ma,Howl:2009ds}, where a mass squared is driven negative at a much higher scale. See also Ref.~\cite{Enqvist:2013eua} for a recent treatment of radiative corrections to inflationary potentials. Such a radiatively corrected potential may be parametrised as~\cite{Varzielas:2006ma,Howl:2009ds},
\begin{equation}
 V(\varphi)\approx C + M^2 \Phi^\dagger \Phi \ln \left( \frac{\Phi^\dagger \Phi}{\Lambda^2} \right) = C + \frac{M^2}{2} \varphi^2  \ln \left( \frac{\varphi^2}{2\Lambda^2} \right).
 \label{eq:inf_pot}
\end{equation}
This potential, schematically depicted in Fig.~\ref{fig:potentials}, generates a vacuum expectation value (VEV) $\langle \varphi \rangle = f=\sqrt{\frac{2}{e}}\Lambda$ for the real scalar field $\varphi$. The constant $C=M^2\,f^2/2$ is chosen in a way to yield a potential with value $V(f)=0$ at its minimum. This corresponds to having a vanishing cosmological constant at the end of inflation. As will be discussed below, the Dark Energy dominating today's Universe is then realised through a quintessence field, one of the key features of the RIDE model. Note that we could in principle have chosen a larger value for $C$, too, which would however complicate the quintessence part. Note further that the predictions for inflation are nearly independent of $C$ as long as $C \ll V(\varphi_N)$, i.e., the constant is significantly smaller than the value of the potential $N$ $e$-folds before the end of inflation.

Interpreting $\varphi$ as the inflaton field, inflation can completely take place in a region where $\varphi \gg \Lambda$. With the $\ln$-factor in Eq.~\eqref{eq:inf_pot} being well-behaved, $\varphi$ feels a potential that is very similar to the one used for quadratic inflation. Therefore, we expect only small differences in the predictions for inflation in the RIDE model compared to $\varphi^2$ chaotic inflation, see Sec.~\ref{sec:Inflation}. 

The main advantage of the RIDE model versus a $\varphi^2$ model of inflation lies in the possibility to incorporate a quintessence field to explain Dark Energy. As the minimum of the RIDE potential, Eq.~\eqref{eq:inf_pot}, is displaced from $\langle \varphi \rangle = 0 $, the $U(1)$ symmetry gets broken when the inflaton field $\varphi$ settles to its VEV. This results in a massless Nambu-Goldstone boson $\phi=f \arg (\Phi)$, corresponding to the phase of the complex scalar field $\Phi$. Gravitational effects~\cite{Rosenfeld:2005mt,Kallosh:1995hi}, called gravitational instantons, then generate a potential for $\phi$ which must respect a discrete shift symmetry $\frac{\phi}{f} \to \frac{\phi}{f} + 2\pi n$, with $n\in \mathbb N$. It is possible to argue that, see~\cite{Sorbo:2008zz}, the so-obtained potential for the quintessence field $\phi$ takes the form
\begin{equation}
 V_q(\phi)=m^4 \left[ 1+\cos \left( \frac{\phi}{f} \right) \right],
 \label{eq:quint_pot}
\end{equation}
where $m$ denotes the Dark Energy scale, determined from the instanton action $S \sim \pi M_P^2/M_{\rm string}^2$ via $m^4=f M_P^3 e^{-S}$~\cite{Kallosh:1995hi,Masso:2006yk,Kaloper:2008fb,SorboTalk} that can easily reproduce the smallness of the observed scale of Dark Energy $\sim 10^{-3}\,{\rm eV}$. As discussed in~\cite{DiBari:2010wg}, the dynamics of the inflaton field $\varphi$ and the quintessence field $\phi$ separate. Therefore we are practically dealing with single-field potentials, so that the RIDE model is expected to be safe from potentially dangerous corrections due to (iso-)curvature fluctuations that can appear in multi-field inflation models~\cite{Lalak:2007vi}.

%%%%%%%%%%%%%%%%%%%%%%%%%%%%%%%%%%%%%%%%%%%%%%%%%%%%%%%%%%%%%%%%%%%%%%
\section{\label{sec:Inflation}RIDE vs.~chaotic  $\boldsymbol{\varphi^2}$ inflation}
%%%%%%%%%%%%%%%%%%%%%%%%%%%%%%%%%%%%%%%%%%%%%%%%%%%%%%%%%%%%%%%%%%%%%%

The RIDE potential of Eq.~\eqref{eq:inf_pot} depends on two mass parameters, $M$ and $\Lambda$. Assuming $\Lambda$ close to the Planck scale, $M_P=1.2\times
10^{19}$ GeV, we define 
\be
\Lambda~=~ \ka\,M_P ,
\ee
where $\ka$ takes a particular value. However it will turn out that the results in what concerns inflation are quite insensitive to values of $\ka$ in the range $0.01-1$. In order to relate the parameters of the potential to physical observables such as the scalar spectral index $n_S$ as well as the tensor-to-scalar ratio $r$, it is convenient to define the slow-roll parameters. There are two common versions of these parameters used in the literature, namely the ``potential slow-roll parameters'' $\epsilon_V,\eta_V$ and the ``Hubble slow-roll parameters''  $\epsilon_H,\eta_H$, which are related as~\cite{Liddle:1994dx} 
\beqn
\epsilon_V=\epsilon_H+ \,\cdots ,\qquad 
\eta_V=\epsilon_H+\eta_H+\,\cdots\,,
\eeqn
where the dots indicate higher-order corrections to these relations. It is typically sufficient to consider the first-order relations, so that one can easily convert from one convention to the other. In the following we adopt the potential slow-roll parameters, which in the RIDE model take the form\footnote{Note that in~\cite{DiBari:2010wg} we chose to work with the Hubble slow-roll parameters instead.}
\beqn
 \eps_V &\equiv &\frac{M_P^2}{16\pi} \left( \frac{V'}{V} \right)^2= \frac{M_P^2}{4\pi \varphi^2} \left(\frac{1+L}{\rho +L} \right)^2\ , 
\label{eq:eps} \\ 
 \eta_V&\equiv&\frac{M_P^2}{8\pi}  \frac{V''}{V}  = \frac{M_P^2}{4\pi  \varphi^2} \frac{3+L}{\rho +L}, 
\label{eq:eta} \\
  \xi_V^2 &\equiv & \frac{M_P^4}{(8\pi)^2} \frac{V' V'''}{V^2} = \frac{M_P^4}{8\pi^2 \varphi^4} \frac{1+ L}{(\rho +L)^2}\ ,
\eeqn
where $\rho = f^2/\varphi^2$ and $L=\ln \left[\varphi^2/(2\Lambda^2) \right]$, and the third derivative potential parameter $\xi_V^2$ will be relevant for the running of the scalar spectral index. Note that these expressions are independent of the parameter $M$.\footnote{Note that $M$ should not be confused with the scale of inflation given by $V^{1/4} \simeq \sqrt{M \varphi_N}/2^{1/4}$.} The end of inflation is reached when the Hubble slow-roll parameter $\epsilon_H=1$. In our numerically calculation of the field value $\varphi_e$ at the end of inflation we make use of the corresponding (approximate) condition $\epsilon_V=1$. In practice, $\varphi_e$ is always well above $f$. Having determined $\varphi_e$, we have to find the field value $\varphi_N$, $N$ $e$-folds before the end of inflation. This is achieved by solving
\begin{equation}
 N~\simeq~\frac{8\pi}{M_P^2} \int_{\varphi_e}^{\varphi_N} \frac{V(\varphi)}{V^\prime(\varphi)} d\varphi ~=~ \frac{2\pi}{M_P^2} \left[ \varphi_N^2 - \varphi_e^2 - f^2 \Big[ {\rm Ei}(1+L_N) - {\rm Ei}(1+L_e) - \ln \Big( \frac{1+L_N}{1+L_e} \Big) \Big] \right],
 \label{eq:eta_N}
\end{equation}
numerically, with ${\rm Ei}(z)$ being the exponential integral ${\rm Ei}(z)=-\int_{-z}^{\infty} \frac{e^{-t}}{t} dt$, $L_i=\ln \left(\frac{\varphi_i^2}{2\Lambda^2} \right)$, and $N$ lying within the interval $N\in [46,60]$. Using the so-obtained value of $\varphi_N$ (for a fixed value of $N$), the parameter $M$ in Eq.~\eqref{eq:inf_pot} can be directly constrained by the observed  scalar perturbations in the Cosmic Microwave Background, as discussed in~\cite{DiBari:2010wg}. One finds $M\simeq [10^{-8} M_P, 10^{-7} M_P]$ where this result is essentially independent of the scale $\Lambda$, thanks to the logarithmic dependence of $V$ on $\Lambda$. 

Starting from certain values for $\ka=\frac{\Lambda}{M_P}$ and $N$, the two parameters of the potential in Eq.~\eqref{eq:inf_pot}, $M$ and $\Lambda$, are thus fixed. Moreover, the field values at the end of inflation $\varphi_e$ as well as $N$ $e$-folds before, $\varphi_N$, are determined via the procedure discussed above. It is now straightforward to calculate the RIDE predictions for the scalar spectral index~$n_S$ as well as the tensor-to-scalar ratio~$r$ for different values of $\ka=\frac{\Lambda}{M_P}$ and $N$ using the first order expressions,
\beqn
n_S&=& 1-6 \eps_V + 2 \eta_V  \ ,\\
r&=&16\eps_V \ .
\eeqn
The running of the scalar spectral index is only important at second order in the small parameters,
\be
\frac{d n_S}{d \ln k} = 16 \epsilon_V \eta_V -24 \epsilon_V^2 -2 \xi_V^2 \  ,
\ee
which is one of the reasons for not taking it into account explicitly in our simple analysis (we will comment on the other reason in a second). For chaotic $\varphi^2$ inflation the expressions for the slow-roll parameters and the relation between the field values $\varphi_N,\varphi_e$ and the number of $e$-foldings $N$  simplify to
\be
\eps_V = \eta_V = \frac{M_P^2}{4\pi \varphi^2} \ , ~~~
N~\simeq~2\pi  \frac{\varphi_N^2 - \varphi_e^2}{M_P^2}  \ ,
\ee
Formally, this can be obtained from Eqs.~(\ref{eq:eps},\ref{eq:eta},\ref{eq:eta_N}) by dropping all terms involving $L$. Evidently $\xi^2_V=0$ in chaotic $\varphi^2$ inflation.

In our statistical study, we have performed a $\chi^2$ analysis of the RIDE model and of the similar chaotic $\varphi^2$ inflation model. In order to do this, we have extracted several ``data'' points from the $1\sigma$ contour in the $(n_S,r)$-plane from the BICEP2 paper~\cite{Ade:2014xna} by fitting a tilted ellipse to it. This fit resulted in a ``best-fit'' point of $(n_{S,\rm best},r_{\rm best}) = (0.959, 0.184)$. Note that this point is \emph{not} identical to the best-fit point given in Ref.~\cite{Ade:2014xna}, which quotes $r = 0.20_{-0.05}^{+0.07}$. This difference simply comes from the fact that the errors in the fitted ellipse are effectively symmetrised (in the basis where the tilt vanishes), but it is numerically not very significant and the resulting error is acceptable in a simplified treatment.

Our simplified fit allows to determine the $2 \times 2$ covariance matrix $F$ and to construct a $\chi^2$ function from it as 
\begin{equation}
 \chi^2 = (n_S - n_{S,\rm best}, r - r_{\rm best}) F
 \begin{pmatrix}
 n_S - n_{S,\rm best}\\
 r - r_{\rm best}
 \end{pmatrix}, \ \ \ {\rm where}\ \ \ F =
 \begin{pmatrix}
 7736.6 & 184.2\\
 184.2 & 211.6
 \end{pmatrix}.
 \label{eq:chi2}
\end{equation}
The numerical values of the matrix $F$ are obtained from the semi-minor and semi-major axes of the fitted ellipse, $a \approx 0.01137$ and $b \approx 0.06948$, by rotating the diagonal matrix ${\rm diag}(\frac{1}{a^2},\frac{1}{b^2})$ by the angle of the tilt ($\approx -1.4^\circ$). Note that we have not taken into account the information on the running in Eq.~\eqref{eq:chi2}, since we have no way to extract the full covariance matrix from Ref.~\cite{Ade:2014xna}, and thus we cannot know how a change in $d n_S/d \ln k$ would affect the other variables. However, if one would nevertheless like to at least approximately take into account the running, one could do this by simply using the corresponding best-fit value and $1\sigma$ range given in Ref.~\cite{Ade:2014xna} and add this information to Eq.~\eqref{eq:chi2} as penalty term, which results in the total $\chi^2$ function:
\begin{equation}
\chi^2_{\rm total} = \chi^2 +  \chi^2_{\rm running} \ , \quad {\rm with}\quad
\chi^2_{\rm running}=  \frac{\left[\frac{d n_S}{d \ln k} - (-0.022)\right]^2}{0.010^2}.
 \label{eq:chi2_running}
\end{equation}
$\chi^2_{\rm total}$ is the function we would then have to minimise. However, we will first focus on $\chi^2$ of Eq.~\eqref{eq:chi2} and postpone a discussion of the contribution from the running to the end of this section.

\begin{figure}[t]
\centering
\includegraphics[width=0.6\textwidth]{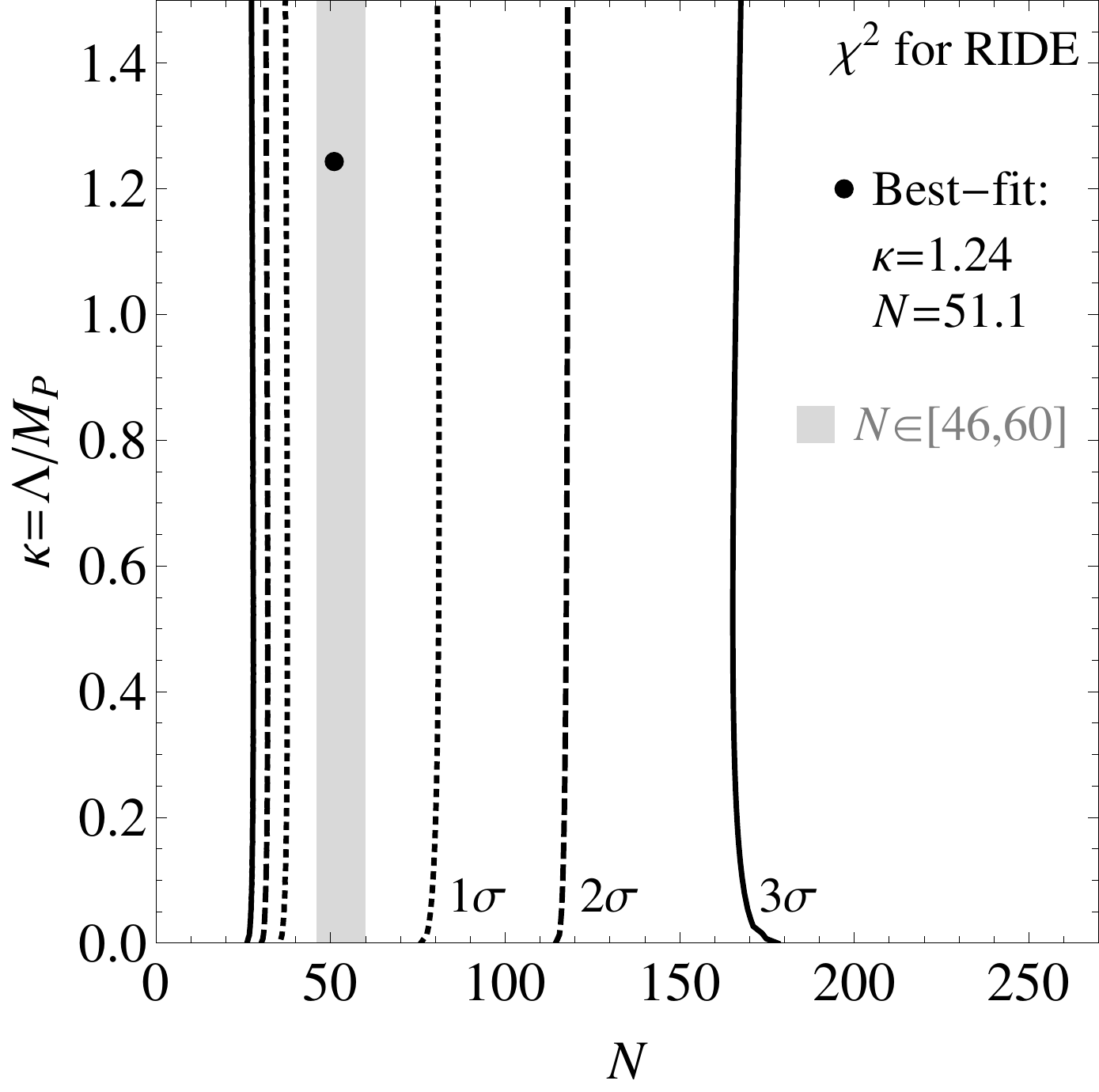}
\vspace*{-4mm}
    \caption{\label{fig:chi2_RIDE}The 2-dimensional $\chi^2$ distribution for the RIDE model parameters. Note that the $\chi^2$ function is nearly flat in the parameter $\ka$, so that the (trans-Planckian) best-fit value $\ka= M_P/\Lambda \simeq 1.24$ is practically arbitrary and the numerical minimisation could have found a local minimum at any value of $\ka$, as long as the number of $e$-folds is $N\simeq 51.1$.}
\vspace*{-2mm}
\end{figure}

Starting with RIDE, which involves the free parameter $\ka= M_P/\Lambda$, we can calculate the $\chi^2$ function in the two-parameter $(N,\ka)$ plane. The result can be seen in Fig.~\ref{fig:chi2_RIDE}, where we have indicated the $1\sigma$ ($2\sigma$, $3\sigma$) region by the dotted (dashed, solid) lines. The best-fit point turns out to be $(N,\ka) = (51.1,1.24)$, with a minimum of $\chi^2 = 0.009$ which signals a nearly perfect fit. However, as visible in the plot, the $\chi^2$ function is nearly flat in $\ka$-direction, so that practically any value of $\ka$ would be allowed [as to be expected since the inflation potential depends only logarithmically on $\ka$, cf.\ Eq.~\eqref{eq:inf_pot}]. Note that quintessence imposes a relatively general bound of $\ka \gtrsim 0.5$~\cite{DiBari:2010wg} on this parameter, if too much fine tuning is to be avoided. We will nevertheless display results also for $\ka = 0.01$ in what follows, in order to make the (weak) dependence of our results on $\ka$ more apparent in the plots. Clearly, the ideal number $N$ of $e$-folds clearly is an important parameter, so that we would expect RIDE to fit best in the vicinity of $N \approx 51$.

\begin{figure}[t]
\centering
\includegraphics[width=0.49\textwidth]{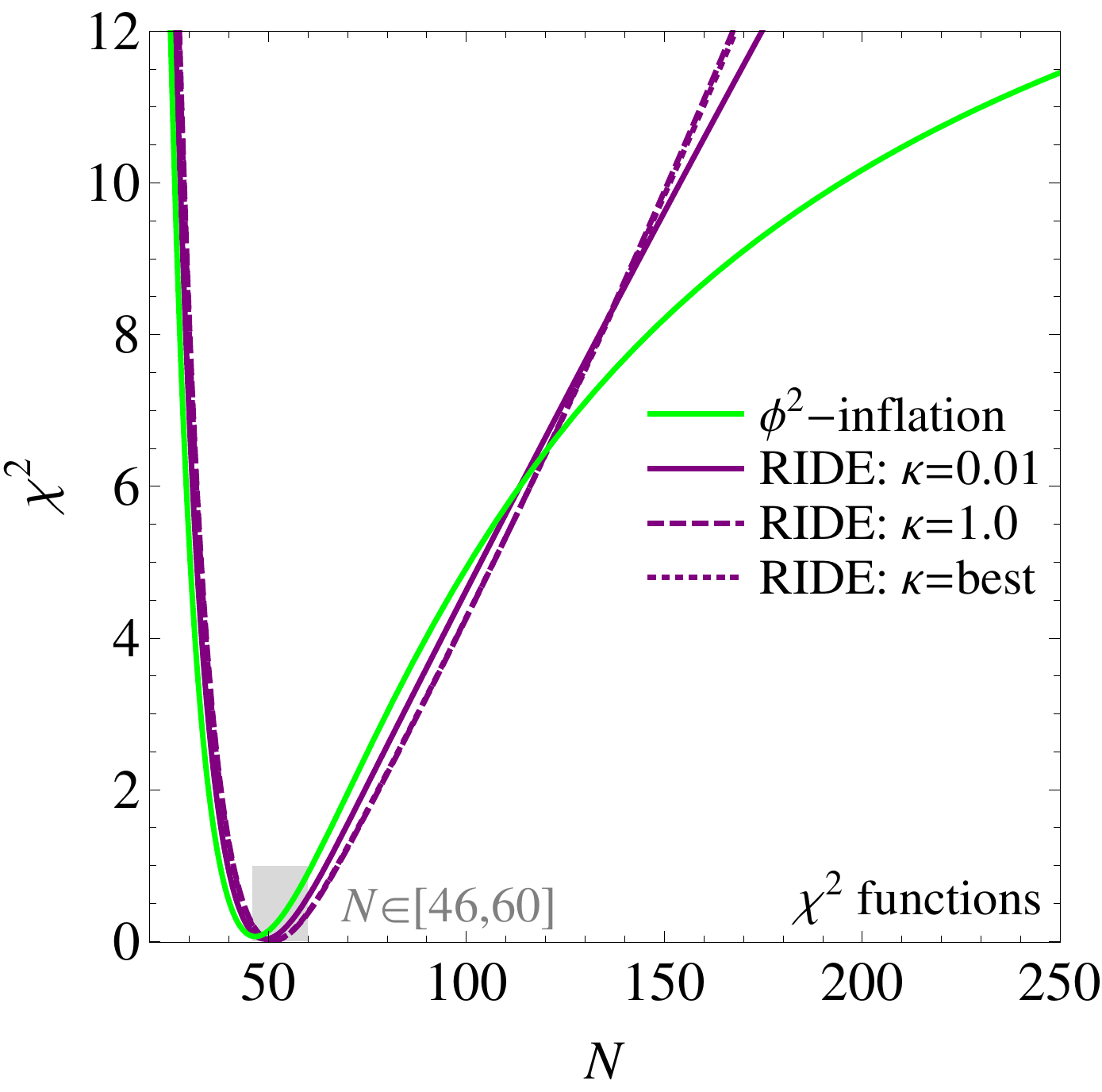}
\includegraphics[width=0.49\textwidth]{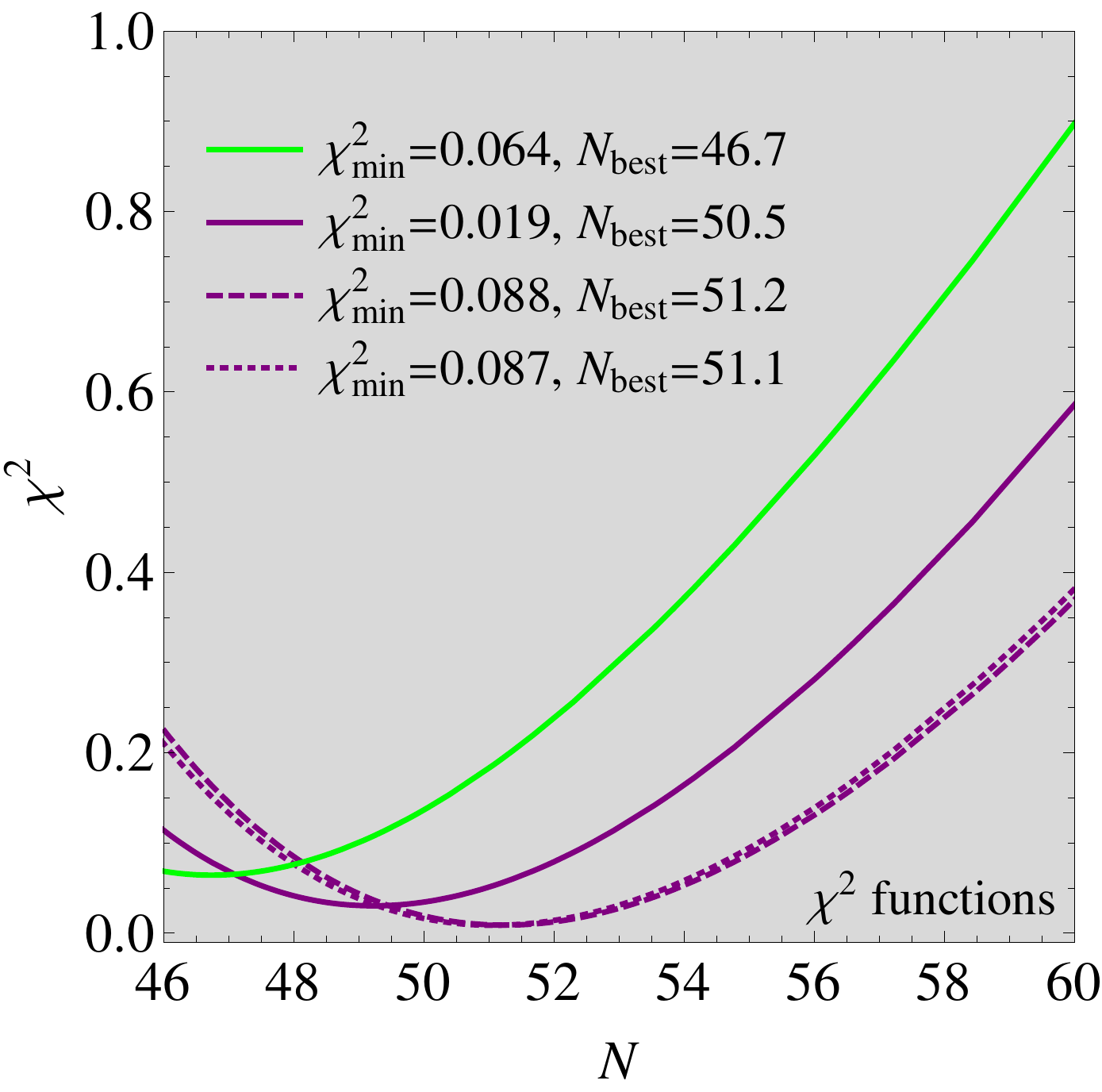}
\vspace*{-4mm}
    \caption{\label{fig:chi2_comparison}The 1-dimensional $\chi^2$ distributions for RIDE (for different values of $\ka=M_P/\Lambda$) 
    and $\varphi^2$ inflation (left panel: larger range for $N$, right panel: $N \in [46,60]$).}
\vspace*{-2mm}
\end{figure}

Thus, we can simply take different values of $\ka$ to find a simple comparison between RIDE and $\varphi^2$ inflation, which is displayed in Fig.~\ref{fig:chi2_comparison}. Glancing at the left panel, it can be seen that all $\chi^2$ functions look pretty similar, at least up to the $2\sigma$ range ($\Delta \chi^2 = 4$), while they do visibly differ at $3\sigma$ ($\Delta \chi^2 = 9$). Indeed when looking closer, see right panel, it is visible that RIDE (for very different values of $\ka$) fits marginally better than $\varphi^2$ inflation. When aiming at distinguishing the two models even only at $1\sigma$ level, one would require an improvement on the knowledge on $(n_S, r)$ by more than about an order of magnitude. A distinction at even higher confidence level (such as $3\sigma$) is unlikely to happen within the foreseeable future for $n_S$. However, it is intriguing that a future CMB polarisation experiment that covers the whole sky and very low instrumental noise might measure $r$ with a precision well below $\Delta r \sim 0.01$, as needed to distinguish RIDE from the $\varphi^2$ model~\cite{Dodelson:2014exa} (and surely even more so from $\varphi^4$ \cite{Okada:2014lxa}). \footnote{Note that, in principle, one could distinguish both models by the \emph{running of the scalar spectral index}, which is reported to be $d n_S/d \ln k = -0.022\pm 0.010$ ($1\sigma$ level)~\cite{Ade:2014xna}, since the third derivative of the RIDE potential does \emph{not} vanish while the one of a $\varphi^2$ potential trivially does. However, in practice both models yield a running which seems too small by about an order of magnitude. We can estimate the running as $\mathcal{O}(10 \epsilon^2)$, which for RIDE is about $0.001$ due to $\epsilon \sim \eta \sim  0.01$, while $\xi^2 \sim 10^{-5}$. A similar size is obtained for ordinary chaotic $\varphi^2$ inflation.}

\begin{figure}[t]
\centering
\includegraphics[width=0.7\textwidth]{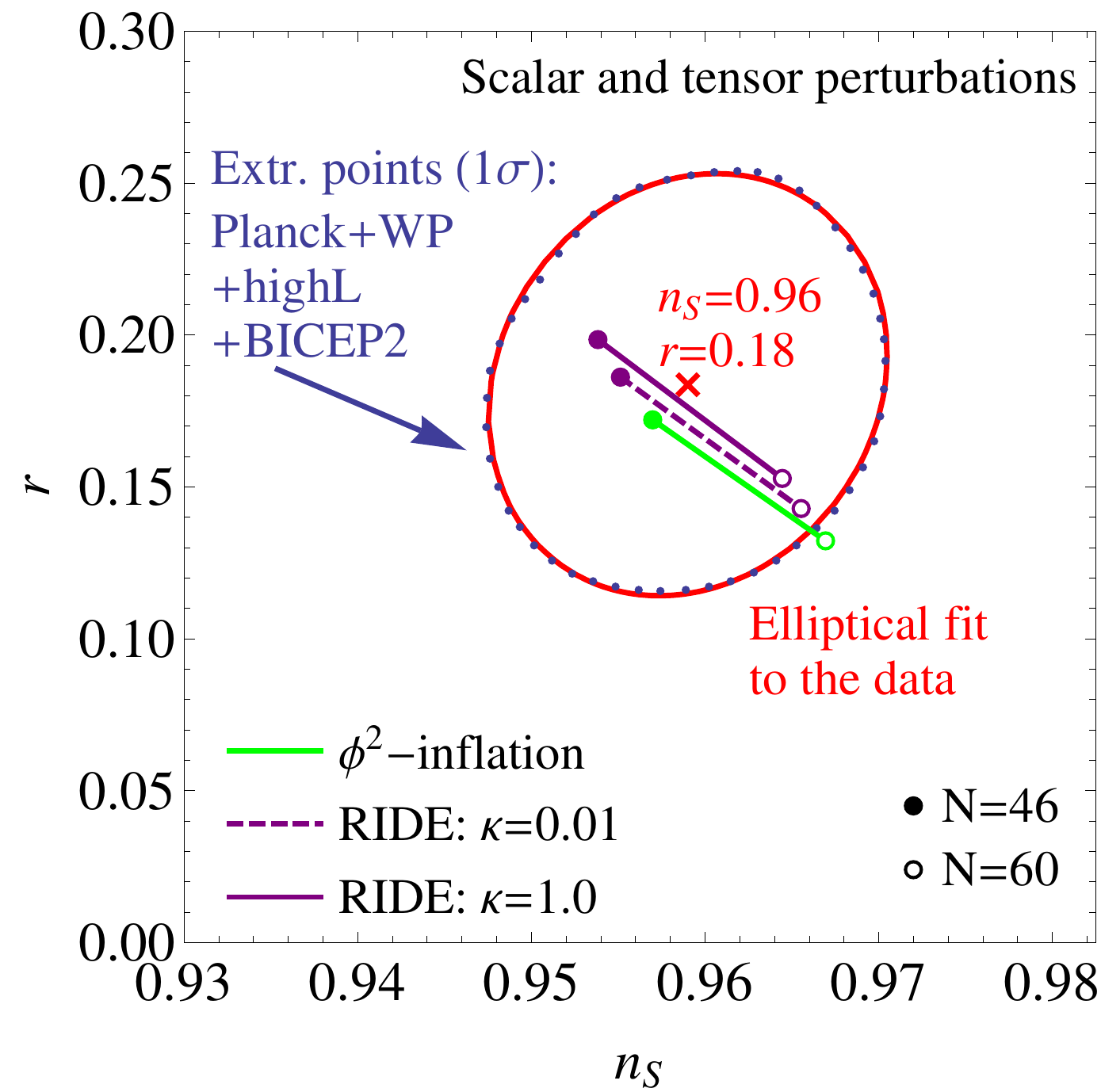}
\vspace*{-4mm}
    \caption{\label{fig:inflation}The predictions of RIDE vs.\ the chaotic $\varphi^2$ inflation model for the spectral index $n_S$ and tensor-to-scalar ratio $r$.}
\vspace*{-2mm}
\end{figure}

We can display the predictions of both models in the $(n_S, r)$-plane, which further indicates their statistical similarity. This is done in Fig.~\ref{fig:inflation}, where also our simple elliptical fit to the extracted ``data'' points is shown. Indeed, both models give predictions within the $1\sigma$ region (except for $\varphi^2$ inflation for values of $N$ close to 60, but even that does not go far out). The detailed predictions for the best-fit values are:
\begin{center}
\begin{tabular}{|c||c|c|c||c|c|}\hline
Model & $\chi^2_{\rm min}$ & $N$ & $\ka$ & $n_S$ & $r$\\ \hline \hline
RIDE  & $0.009$ & $51.1$ & $1.24$ & $0.958$ & $0.179$\\ \hline
$\varphi^2$ inflation  & $0.064$ & $46.7$ & --- & $0.958$ & $0.169$\\ \hline
\end{tabular}
\end{center}
Again we can see that the predictions are very similar, if not nearly identical. 

Digesting these results, one could question whether any of the two models has any advantage. Indeed there is at least a ``theoretical advantage'' to the RIDE model: as described in Ref.~\cite{DiBari:2010wg}, this model also includes the possibility to explain Dark Energy in terms of a cosine quintessence potential. But, more realistically, it is simply a viable alternative to $\varphi^2$ inflation, and thus one of only a few \emph{working} examples combining inflation with Dark Energy\footnote{It should be noticed that BICEP2 resurrected $\varphi^4$ models as well, though more marginally \cite{Okada:2014lxa}.}. If the parameter space shrinks around those two models, both of them would probably have to be viewed as more or less equally good competitors for the true theory behind inflation, since at the moment they are not distinguishable on a statistical basis, as we have shown.

The results on $(n_S,r)$ by combining the BICEP2 results with those from {\em Planck} temperature anisotropies data, WMAP polarisation, and high${\ell}$ experiments as shown in Fig.~4 are obtained allowing for a non-vanishing running of the scalar spectral index and find $d n_S/d \ln k = -0.022\pm 0.010$. As mentioned already, without running BICEP2 results always give $r\simeq 0.2$, while the {\em Planck}+WP+high${\ell}$ data  lead to an upper bound $r < 0.11 (95\%~\rm{C.L.})$ in tension with BICEP2 results. This seems to strengthen the already existing hint for a non-vanishing value of $d n_S/d \ln k \sim - 0.01$ both from WMAP7~\cite{Komatsu:2010fb} and {\em Planck}~\cite{Ade:2013zuv} temperature anisotropies data. The best-fit for the absolute value would be too large compared to the values  from single-field inflationary models, including RIDE, predicting $|d n_S/d\ln k |\sim (n_S -1)^2 \sim 1/N^2 \sim 10^{-3}$.  Including errors, the contribution from running to $\chi_{\rm total}^2$ in Eq.~(\ref{eq:chi2_running}) gives $\chi_{\rm total}^2 - \chi^2 \sim 4$,  quantifying the $\sim 2\sigma$ tension. It should be said, however, that such a tension might be resolved by a better account of foreground contamination or by other systematics. Future data, in particular those expected  from {\em Planck} on $B$-mode polarisation, will be able to strongly reduce the systematic uncertainties and lead to a much higher statistically significant measurement of the running. Therefore, they will be a crucial test for RIDE, as for all other simple single field inflationary models, that might be (more strongly supported) ruled out if large values of $|dn_S/d\ln k| \sim 10^{-2}$ will (not) be  confirmed.

%%%%%%%%%%%%%%%%%%%%%%%%%%%%%%%%%%%%%%%%%%%%%%%%%%%%%%%%%%%%%%%%%%%%%%
\section{\label{sec:Conclusions}Conclusions}
%%%%%%%%%%%%%%%%%%%%%%%%%%%%%%%%%%%%%%%%%%%%%%%%%%%%%%%%%%%%%%%%%%%%%%

We have revisited the RIDE model based on radiative symmetry breaking that combines inflation with Dark Energy. We have performed a $\chi^2$ analysis for the RIDE model parameters and have compared the predictions of RIDE vs.\ the chaotic $\varphi^2$ inflation model for the spectral index $n_S$ and tensor-to-scalar ratio $r$. The RIDE model gives a slightly better fit to the data than the chaotic $\varphi^2$ inflation model. To be precise we find a best-fit value in the RIDE model of $r=0.18$ as compared to $r=0.17$ in the chaotic model, with the spectral index being $n_S=0.96$ in both models. In addition, RIDE has the \emph{additional advantage} that it accounts for the Dark Energy of the universe via the PNGB quintessence field generated at the end of inflation.

%%%%%%%%%%%%%%%%%%%%%%%%%%%%%%%%%%%%%%%%%%%%%%%%%%%%%%%%%%%%%%%%%%%%%%
\section*{\label{sec:Note Added}Note added}
%%%%%%%%%%%%%%%%%%%%%%%%%%%%%%%%%%%%%%%%%%%%%%%%%%%%%%%%%%%%%%%%%%%%%%

After completion of our work, Refs.~\cite{Mortonson:2014bja,Flauger:2014qra}
appeared where the  authors point out how the $B$-mode polarisation signal
reported by the BICEP2 collaboration could be partially, if not entirely, due
to a foreground contamination effect. In light of this possible effect, 
much lower values of $r$, even a vanishing one, cannot be currently excluded
and new results from {\em Planck} or ground based telescopes, such as future
Keck Array observations, measuring the signal at different frequencies are
needed in order to resolve this ambiguity. As discussed above, our RIDE model 
fits very well with the BICEP2 determination of $r\simeq 0.2$, and gives a
best fit value of $r\simeq 0.18$. As can be inferred from Fig.~3, 
a determination of $r\sim 0.1$, at the level of the current $95\%\,{\rm C.L.}$
{\em Planck} upper bound, with an error $\Delta r \simeq 0.05$ as for BICEP2,
would start to be in tension with the RIDE model, but still
compatible. Measuring $r\sim 0.1$ with a smaller error $\Delta r\sim 0.01$, as
expected from future observations, would, however, strongly disfavour our RIDE model.

%%%%%%%%%%%%%%%%%%%%%%%%%%%%%%%%%%%%%%%%%%%%%%%%%%%%%%%%%%%%%%%%%%%%%%
\section*{\label{sec:Ack}\noindent{Acknowledgments}}
%%%%%%%%%%%%%%%%%%%%%%%%%%%%%%%%%%%%%%%%%%%%%%%%%%%%%%%%%%%%%%%%%%%%%%

PDB acknowledges financial support from the NExT/SEPnet Institute and from the STFC Rolling Grant ST/G000557/1. SFK acknowledges support from the STFC Consolidated ST/J000396/1 grant. The work of CL is supported by the Deutsche Forschungsgemeinschaft (DFG) within the Research Unit FOR 1873 (Quark Flavour Physics and Effective Field Theories). AM acknowledges support by a Marie Curie Intra-European Fellowship within the 7th European Community Framework Programme FP7-PEOPLE-2011-IEF, contract PIEF-GA-2011-297557. Finally PDB, SFK, and AM all three acknowledge partial support from the European Union FP7 ITN-INVISIBLES (Marie Curie Actions, PITN-GA-2011-289442).

%=============================================================================
\bibliographystyle{./apsrev}
\bibliography{./RIDE}

\begin{thebibliography}{10}
\expandafter\ifx\csname bibnamefont\endcsname\relax
  \def\bibnamefont#1{#1}\fi
\expandafter\ifx\csname bibfnamefont\endcsname\relax
  \def\bibfnamefont#1{#1}\fi
\expandafter\ifx\csname url\endcsname\relax
  \def\url#1{\texttt{#1}}\fi
\expandafter\ifx\csname urlprefix\endcsname\relax\def\urlprefix{URL }\fi
\providecommand{\bibinfo}[2]{#2}
\providecommand{\eprint}[2][]{\url{#2}}

\bibitem{Ade:2014xna}
\bibinfo{author}{\bibfnamefont{P.~A.~R.} \bibnamefont{Ade}} \emph{et~al.}
  (\bibinfo{collaboration}{BICEP2 Collaboration})  (\bibinfo{year}{2014}),
  \eprint{1403.3985}.

\bibitem{Lyth:1998xn}
\bibinfo{author}{\bibfnamefont{D.~H.} \bibnamefont{Lyth}} \bibnamefont{and}
  \bibinfo{author}{\bibfnamefont{A.}~\bibnamefont{Riotto}},
  \bibinfo{journal}{Phys. Rept.} \textbf{\bibinfo{volume}{314}},
  \bibinfo{pages}{1} (\bibinfo{year}{1999}), \eprint{hep-ph/9807278}.

\bibitem{Ade:2013zuv}
\bibinfo{author}{\bibfnamefont{P.}~\bibnamefont{Ade}} \emph{et~al.}
  (\bibinfo{collaboration}{Planck Collaboration})  (\bibinfo{year}{2013}),
  \eprint{1303.5076}.

\bibitem{Harigaya:2014sua}
\bibinfo{author}{\bibfnamefont{K.}~\bibnamefont{Harigaya}},
  \bibinfo{author}{\bibfnamefont{M.}~\bibnamefont{Ibe}},
  \bibinfo{author}{\bibfnamefont{K.}~\bibnamefont{Schmitz}}, \bibnamefont{and}
  \bibinfo{author}{\bibfnamefont{T.~T.} \bibnamefont{Yanagida}}
  (\bibinfo{year}{2014}), \eprint{1403.4536}.

\bibitem{Harigaya:2014qza}
\bibinfo{author}{\bibfnamefont{K.}~\bibnamefont{Harigaya}} \bibnamefont{and}
  \bibinfo{author}{\bibfnamefont{T.~T.} \bibnamefont{Yanagida}}
  (\bibinfo{year}{2014}), \eprint{1403.4729}.

\bibitem{Lee:2014spa}
\bibinfo{author}{\bibfnamefont{H.~M.} \bibnamefont{Lee}}
  (\bibinfo{year}{2014}), \eprint{1403.5602}.

\bibitem{Hazra:2014aea}
\bibinfo{author}{\bibfnamefont{D.~K.} \bibnamefont{Hazra}},
  \bibinfo{author}{\bibfnamefont{A.}~\bibnamefont{Shafieloo}},
  \bibinfo{author}{\bibfnamefont{G.~F.} \bibnamefont{Smoot}}, \bibnamefont{and}
  \bibinfo{author}{\bibfnamefont{A.~A.} \bibnamefont{Starobinsky}}
  (\bibinfo{year}{2014}), \eprint{1403.7786}.

\bibitem{Channuie:2013xoa}
\bibinfo{author}{\bibfnamefont{P.}~\bibnamefont{Channuie}}
  (\bibinfo{year}{2013}), \eprint{1312.7122}.

\bibitem{Joergensen:2014rya}
\bibinfo{author}{\bibfnamefont{J.}~\bibnamefont{Joergensen}},
  \bibinfo{author}{\bibfnamefont{F.}~\bibnamefont{Sannino}}, \bibnamefont{and}
  \bibinfo{author}{\bibfnamefont{O.}~\bibnamefont{Svendsen}}
  (\bibinfo{year}{2014}), \eprint{1403.3289}.

\bibitem{Pallis:2014dma}
\bibinfo{author}{\bibfnamefont{C.}~\bibnamefont{Pallis}}
  (\bibinfo{year}{2014}), \eprint{1403.5486}.

\bibitem{Bamba:2014jia}
\bibinfo{author}{\bibfnamefont{K.}~\bibnamefont{Bamba}},
  \bibinfo{author}{\bibfnamefont{R.}~\bibnamefont{Myrzakulov}},
  \bibinfo{author}{\bibfnamefont{S.~D.} \bibnamefont{Odintsov}},
  \bibnamefont{and}
  \bibinfo{author}{\bibfnamefont{L.}~\bibnamefont{Sebastiani}}
  (\bibinfo{year}{2014}), \eprint{1403.6649}.

\bibitem{Bezrukov:2014nza}
\bibinfo{author}{\bibfnamefont{F.}~\bibnamefont{Bezrukov}} \bibnamefont{and}
  \bibinfo{author}{\bibfnamefont{D.}~\bibnamefont{Gorbunov}}
  (\bibinfo{year}{2014}), \eprint{1403.4638}.

\bibitem{Ferrara:2014ima}
\bibinfo{author}{\bibfnamefont{S.}~\bibnamefont{Ferrara}},
  \bibinfo{author}{\bibfnamefont{A.}~\bibnamefont{Kehagias}}, \bibnamefont{and}
  \bibinfo{author}{\bibfnamefont{A.}~\bibnamefont{Riotto}}
  (\bibinfo{year}{2014}), \eprint{1403.5531}.

\bibitem{Kidani:2014pka}
\bibinfo{author}{\bibfnamefont{T.}~\bibnamefont{Kidani}} \bibnamefont{and}
  \bibinfo{author}{\bibfnamefont{K.}~\bibnamefont{Koyama}}
  (\bibinfo{year}{2014}), \eprint{1403.6687}.

\bibitem{Kallosh:2014qta}
\bibinfo{author}{\bibfnamefont{R.}~\bibnamefont{Kallosh}},
  \bibinfo{author}{\bibfnamefont{A.}~\bibnamefont{Linde}},
  \bibinfo{author}{\bibfnamefont{B.}~\bibnamefont{Vercnocke}},
  \bibnamefont{and}
  \bibinfo{author}{\bibfnamefont{W.}~\bibnamefont{Chemissany}}
  (\bibinfo{year}{2014}), \eprint{1403.7189}.

\bibitem{Hamaguchi:2014mza}
\bibinfo{author}{\bibfnamefont{K.}~\bibnamefont{Hamaguchi}},
  \bibinfo{author}{\bibfnamefont{T.}~\bibnamefont{Moroi}}, \bibnamefont{and}
  \bibinfo{author}{\bibfnamefont{T.}~\bibnamefont{Terada}}
  (\bibinfo{year}{2014}), \eprint{1403.7521}.

\bibitem{Nakayama:2014koa}
\bibinfo{author}{\bibfnamefont{K.}~\bibnamefont{Nakayama}} \bibnamefont{and}
  \bibinfo{author}{\bibfnamefont{F.}~\bibnamefont{Takahashi}}
  (\bibinfo{year}{2014}), \eprint{1403.4132}.

\bibitem{Cook:2014dga}
\bibinfo{author}{\bibfnamefont{J.~L.} \bibnamefont{Cook}},
  \bibinfo{author}{\bibfnamefont{L.~M.} \bibnamefont{Krauss}},
  \bibinfo{author}{\bibfnamefont{A.~J.} \bibnamefont{Long}}, \bibnamefont{and}
  \bibinfo{author}{\bibfnamefont{S.}~\bibnamefont{Sabharwal}}
  (\bibinfo{year}{2014}), \eprint{1403.4971}.

\bibitem{Germani:2014hqa}
\bibinfo{author}{\bibfnamefont{C.}~\bibnamefont{Germani}},
  \bibinfo{author}{\bibfnamefont{Y.}~\bibnamefont{Watanabe}}, \bibnamefont{and}
  \bibinfo{author}{\bibfnamefont{N.}~\bibnamefont{Wintergerst}}
  (\bibinfo{year}{2014}), \eprint{1403.5766}.

\bibitem{Bezrukov:2014bra}
\bibinfo{author}{\bibfnamefont{F.}~\bibnamefont{Bezrukov}} \bibnamefont{and}
  \bibinfo{author}{\bibfnamefont{M.}~\bibnamefont{Shaposhnikov}}
  (\bibinfo{year}{2014}), \eprint{1403.6078}.

\bibitem{Costa:2014lta}
\bibinfo{author}{\bibfnamefont{R.}~\bibnamefont{Costa}} \bibnamefont{and}
  \bibinfo{author}{\bibfnamefont{H.}~\bibnamefont{Nastase}}
  (\bibinfo{year}{2014}), \eprint{1403.7157}.

\bibitem{Fairbairn:2014nxa}
\bibinfo{author}{\bibfnamefont{M.}~\bibnamefont{Fairbairn}},
  \bibinfo{author}{\bibfnamefont{P.}~\bibnamefont{Grothaus}}, \bibnamefont{and}
  \bibinfo{author}{\bibfnamefont{R.}~\bibnamefont{Hogan}}
  (\bibinfo{year}{2014}), \eprint{1403.7483}.

\bibitem{Hamada:2014iga}
\bibinfo{author}{\bibfnamefont{Y.}~\bibnamefont{Hamada}},
  \bibinfo{author}{\bibfnamefont{H.}~\bibnamefont{Kawai}},
  \bibinfo{author}{\bibfnamefont{K.-y.} \bibnamefont{Oda}}, \bibnamefont{and}
  \bibinfo{author}{\bibfnamefont{S.~C.} \bibnamefont{Park}}
  (\bibinfo{year}{2014}), \eprint{1403.5043}.

\bibitem{Dimopoulos:2014boa}
\bibinfo{author}{\bibfnamefont{K.}~\bibnamefont{Dimopoulos}}
  (\bibinfo{year}{2014}), \eprint{1403.4071}.

\bibitem{Craig:2014rta}
\bibinfo{author}{\bibfnamefont{N.}~\bibnamefont{Craig}} \bibnamefont{and}
  \bibinfo{author}{\bibfnamefont{D.}~\bibnamefont{Green}}
  (\bibinfo{year}{2014}), \eprint{1403.7193}.

\bibitem{Ellis:2014rxa}
\bibinfo{author}{\bibfnamefont{J.}~\bibnamefont{Ellis}},
  \bibinfo{author}{\bibfnamefont{M.~A.~G.} \bibnamefont{Garcia}},
  \bibinfo{author}{\bibfnamefont{D.~V.} \bibnamefont{Nanopoulos}},
  \bibnamefont{and} \bibinfo{author}{\bibfnamefont{K.~A.} \bibnamefont{Olive}}
  (\bibinfo{year}{2014}), \eprint{1403.7518}.

\bibitem{Lyth:2014yya}
\bibinfo{author}{\bibfnamefont{D.~H.} \bibnamefont{Lyth}}
  (\bibinfo{year}{2014}), \eprint{1403.7323}.

\bibitem{Byrnes:2014xua}
\bibinfo{author}{\bibfnamefont{C.~T.} \bibnamefont{Byrnes}},
  \bibinfo{author}{\bibfnamefont{M.}~\bibnamefont{Cortês}}, \bibnamefont{and}
  \bibinfo{author}{\bibfnamefont{A.~R.} \bibnamefont{Liddle}}
  (\bibinfo{year}{2014}), \eprint{1403.4591}.

\bibitem{Freese:2014nla}
\bibinfo{author}{\bibfnamefont{K.}~\bibnamefont{Freese}} \bibnamefont{and}
  \bibinfo{author}{\bibfnamefont{W.~H.} \bibnamefont{Kinney}}
  (\bibinfo{year}{2014}), \eprint{1403.5277}.

\bibitem{Czerny:2014qqa}
\bibinfo{author}{\bibfnamefont{M.}~\bibnamefont{Czerny}},
  \bibinfo{author}{\bibfnamefont{T.}~\bibnamefont{Higaki}}, \bibnamefont{and}
  \bibinfo{author}{\bibfnamefont{F.}~\bibnamefont{Takahashi}}
  (\bibinfo{year}{2014}), \eprint{1403.5883}.

\bibitem{Kobayashi:2014jga}
\bibinfo{author}{\bibfnamefont{T.}~\bibnamefont{Kobayashi}} \bibnamefont{and}
  \bibinfo{author}{\bibfnamefont{O.}~\bibnamefont{Seto}}
  (\bibinfo{year}{2014}), \eprint{1403.5055}.

\bibitem{Okada:2014lxa}
\bibinfo{author}{\bibfnamefont{N.}~\bibnamefont{Okada}},
  \bibinfo{author}{\bibfnamefont{V.~N.} \bibnamefont{\c{S}eno\u{g}uz}},
  \bibnamefont{and} \bibinfo{author}{\bibfnamefont{Q.}~\bibnamefont{Shafi}}
  (\bibinfo{year}{2014}), \eprint{1403.6403}.

\bibitem{Choudhury:2014kma}
\bibinfo{author}{\bibfnamefont{S.}~\bibnamefont{Choudhury}} \bibnamefont{and}
  \bibinfo{author}{\bibfnamefont{A.}~\bibnamefont{Mazumdar}}
  (\bibinfo{year}{2014}), \eprint{1403.5549}.

\bibitem{Choudhury:2013iaa}
\bibinfo{author}{\bibfnamefont{S.}~\bibnamefont{Choudhury}} \bibnamefont{and}
  \bibinfo{author}{\bibfnamefont{A.}~\bibnamefont{Mazumdar}},
  \bibinfo{journal}{Nucl.Phys.} \textbf{\bibinfo{volume}{B882}},
  \bibinfo{pages}{386} (\bibinfo{year}{2014}), \eprint{1306.4496}.

\bibitem{Hotchkiss:2011gz}
\bibinfo{author}{\bibfnamefont{S.}~\bibnamefont{Hotchkiss}},
  \bibinfo{author}{\bibfnamefont{A.}~\bibnamefont{Mazumdar}}, \bibnamefont{and}
  \bibinfo{author}{\bibfnamefont{S.}~\bibnamefont{Nadathur}},
  \bibinfo{journal}{JCAP} \textbf{\bibinfo{volume}{1202}}, \bibinfo{pages}{008}
  (\bibinfo{year}{2012}), \eprint{1110.5389}.

\bibitem{Gong:2014cqa}
\bibinfo{author}{\bibfnamefont{Y.}~\bibnamefont{Gong}}  (\bibinfo{year}{2014}),
  \eprint{1403.5716}.

\bibitem{McDonald:2014kia}
\bibinfo{author}{\bibfnamefont{J.}~\bibnamefont{McDonald}}
  (\bibinfo{year}{2014}), \eprint{1403.6650}.

\bibitem{Zhang:2014dxk}
\bibinfo{author}{\bibfnamefont{J.-F.} \bibnamefont{Zhang}},
  \bibinfo{author}{\bibfnamefont{Y.-H.} \bibnamefont{Li}}, \bibnamefont{and}
  \bibinfo{author}{\bibfnamefont{X.}~\bibnamefont{Zhang}}
  (\bibinfo{year}{2014}), \eprint{1403.7028}.

\bibitem{Dodelson:2014exa}
\bibinfo{author}{\bibfnamefont{S.}~\bibnamefont{Dodelson}}
  (\bibinfo{year}{2014}), \eprint{1403.6310}.

\bibitem{Cheng:2014ota}
\bibinfo{author}{\bibfnamefont{C.}~\bibnamefont{Cheng}} \bibnamefont{and}
  \bibinfo{author}{\bibfnamefont{Q.-G.} \bibnamefont{Huang}}
  (\bibinfo{year}{2014}), \eprint{1403.7173}.

\bibitem{Cheng:2014bma}
\bibinfo{author}{\bibfnamefont{C.}~\bibnamefont{Cheng}} \bibnamefont{and}
  \bibinfo{author}{\bibfnamefont{Q.-G.} \bibnamefont{Huang}}
  (\bibinfo{year}{2014}), \eprint{1403.5463}.

\bibitem{Kinney:2009vz}
\bibinfo{author}{\bibfnamefont{W.~H.} \bibnamefont{Kinney}}
  (\bibinfo{year}{2009}), \eprint{0902.1529}.

\bibitem{Copeland:2006wr}
\bibinfo{author}{\bibfnamefont{E.~J.} \bibnamefont{Copeland}},
  \bibinfo{author}{\bibfnamefont{M.}~\bibnamefont{Sami}}, \bibnamefont{and}
  \bibinfo{author}{\bibfnamefont{S.}~\bibnamefont{Tsujikawa}},
  \bibinfo{journal}{Int. J. Mod. Phys.} \textbf{\bibinfo{volume}{D15}},
  \bibinfo{pages}{1753} (\bibinfo{year}{2006}), \eprint{hep-th/0603057}.

\bibitem{DiBari:2010wg}
\bibinfo{author}{\bibfnamefont{P.}~\bibnamefont{Di~Bari}},
  \bibinfo{author}{\bibfnamefont{S.~F.} \bibnamefont{King}},
  \bibinfo{author}{\bibfnamefont{C.}~\bibnamefont{Luhn}},
  \bibinfo{author}{\bibfnamefont{A.}~\bibnamefont{Merle}}, \bibnamefont{and}
  \bibinfo{author}{\bibfnamefont{A.}~\bibnamefont{Schmidt-May}},
  \bibinfo{journal}{Phys. Rev.} \textbf{\bibinfo{volume}{D84}},
  \bibinfo{pages}{083524} (\bibinfo{year}{2011}), \eprint{1010.5729}.

\bibitem{Rosenfeld:2005mt}
\bibinfo{author}{\bibfnamefont{R.}~\bibnamefont{Rosenfeld}} \bibnamefont{and}
  \bibinfo{author}{\bibfnamefont{J.~A.} \bibnamefont{Frieman}},
  \bibinfo{journal}{JCAP} \textbf{\bibinfo{volume}{0509}}, \bibinfo{pages}{003}
  (\bibinfo{year}{2005}), \eprint{astro-ph/0504191}.

\bibitem{Hill:1988bu}
\bibinfo{author}{\bibfnamefont{C.~T.} \bibnamefont{Hill}} \bibnamefont{and}
  \bibinfo{author}{\bibfnamefont{G.~G.} \bibnamefont{Ross}},
  \bibinfo{journal}{Nucl. Phys.} \textbf{\bibinfo{volume}{B311}},
  \bibinfo{pages}{253} (\bibinfo{year}{1988}).

\bibitem{Frieman:1991tu}
\bibinfo{author}{\bibfnamefont{J.~A.} \bibnamefont{Frieman}},
  \bibinfo{author}{\bibfnamefont{C.~T.} \bibnamefont{Hill}}, \bibnamefont{and}
  \bibinfo{author}{\bibfnamefont{R.}~\bibnamefont{Watkins}},
  \bibinfo{journal}{Phys. Rev.} \textbf{\bibinfo{volume}{D46}},
  \bibinfo{pages}{1226} (\bibinfo{year}{1992}).

\bibitem{Frieman:1995pm}
\bibinfo{author}{\bibfnamefont{J.~A.} \bibnamefont{Frieman}},
  \bibinfo{author}{\bibfnamefont{C.~T.} \bibnamefont{Hill}},
  \bibinfo{author}{\bibfnamefont{A.}~\bibnamefont{Stebbins}}, \bibnamefont{and}
  \bibinfo{author}{\bibfnamefont{I.}~\bibnamefont{Waga}},
  \bibinfo{journal}{Phys. Rev. Lett.} \textbf{\bibinfo{volume}{75}},
  \bibinfo{pages}{2077} (\bibinfo{year}{1995}), \eprint{astro-ph/9505060}.

\bibitem{Hinshaw:2008kr}
\bibinfo{author}{\bibfnamefont{G.}~\bibnamefont{Hinshaw}} \emph{et~al.}
  (\bibinfo{collaboration}{WMAP}), \bibinfo{journal}{Astrophys. J. Suppl.}
  \textbf{\bibinfo{volume}{180}}, \bibinfo{pages}{225} (\bibinfo{year}{2009}),
  \eprint{0803.0732}.

\bibitem{Kallosh:1995hi}
\bibinfo{author}{\bibfnamefont{R.}~\bibnamefont{Kallosh}},
  \bibinfo{author}{\bibfnamefont{A.~D.} \bibnamefont{Linde}},
  \bibinfo{author}{\bibfnamefont{D.~A.} \bibnamefont{Linde}}, \bibnamefont{and}
  \bibinfo{author}{\bibfnamefont{L.}~\bibnamefont{Susskind}},
  \bibinfo{journal}{Phys. Rev.} \textbf{\bibinfo{volume}{D52}},
  \bibinfo{pages}{912} (\bibinfo{year}{1995}), \eprint{hep-th/9502069}.

\bibitem{Martin:2013tda}
\bibinfo{author}{\bibfnamefont{J.}~\bibnamefont{Martin}},
  \bibinfo{author}{\bibfnamefont{C.}~\bibnamefont{Ringeval}}, \bibnamefont{and}
  \bibinfo{author}{\bibfnamefont{V.}~\bibnamefont{Vennin}}
  (\bibinfo{year}{2013}), \eprint{1303.3787}.

\bibitem{Komatsu:2010fb}
\bibinfo{author}{\bibfnamefont{E.}~\bibnamefont{Komatsu}} \emph{et~al.}
  (\bibinfo{year}{2010}), \eprint{1001.4538}.

\bibitem{Ibanez:1982fr}
\bibinfo{author}{\bibfnamefont{L.~E.} \bibnamefont{Ibanez}} \bibnamefont{and}
  \bibinfo{author}{\bibfnamefont{G.~G.} \bibnamefont{Ross}},
  \bibinfo{journal}{Phys. Lett.} \textbf{\bibinfo{volume}{B110}},
  \bibinfo{pages}{215} (\bibinfo{year}{1982}).

\bibitem{Varzielas:2006ma}
\bibinfo{author}{\bibfnamefont{I.}~\bibnamefont{de~Medeiros~Varzielas}}
  \bibnamefont{and} \bibinfo{author}{\bibfnamefont{G.~G.} \bibnamefont{Ross}}
  (\bibinfo{year}{2006}), \eprint{hep-ph/0612220}.

\bibitem{Howl:2009ds}
\bibinfo{author}{\bibfnamefont{R.}~\bibnamefont{Howl}} \bibnamefont{and}
  \bibinfo{author}{\bibfnamefont{S.~F.} \bibnamefont{King}},
  \bibinfo{journal}{Phys. Lett.} \textbf{\bibinfo{volume}{B687}},
  \bibinfo{pages}{355} (\bibinfo{year}{2010}), \eprint{0908.2067}.

\bibitem{Enqvist:2013eua}
\bibinfo{author}{\bibfnamefont{K.}~\bibnamefont{Enqvist}} \bibnamefont{and}
  \bibinfo{author}{\bibfnamefont{M.}~\bibnamefont{Karciauskas}},
  \bibinfo{journal}{JCAP} \textbf{\bibinfo{volume}{1402}}, \bibinfo{pages}{034}
  (\bibinfo{year}{2014}), \eprint{1312.5944}.

\bibitem{Sorbo:2008zz}
\bibinfo{author}{\bibfnamefont{L.}~\bibnamefont{Sorbo}}, \bibinfo{journal}{Mod.
  Phys. Lett.} \textbf{\bibinfo{volume}{A23}}, \bibinfo{pages}{979}
  (\bibinfo{year}{2008}).

\bibitem{Masso:2006yk}
\bibinfo{author}{\bibfnamefont{E.}~\bibnamefont{Masso}} \bibnamefont{and}
  \bibinfo{author}{\bibfnamefont{G.}~\bibnamefont{Zsembinszki}},
  \bibinfo{journal}{JCAP} \textbf{\bibinfo{volume}{0602}}, \bibinfo{pages}{012}
  (\bibinfo{year}{2006}), \eprint{astro-ph/0602166}.

\bibitem{Kaloper:2008fb}
\bibinfo{author}{\bibfnamefont{N.}~\bibnamefont{Kaloper}} \bibnamefont{and}
  \bibinfo{author}{\bibfnamefont{L.}~\bibnamefont{Sorbo}},
  \bibinfo{journal}{Phys. Rev. Lett.} \textbf{\bibinfo{volume}{102}},
  \bibinfo{pages}{121301} (\bibinfo{year}{2009}), \eprint{0811.1989}.

\bibitem{SorboTalk}
\bibinfo{author}{\bibfnamefont{L.}~\bibnamefont{Sorbo}}
  \bibinfo{note}{\emph{Invited talk at the Theory Colloquium, INFN Padua,
  Italy, Mar. 05, 2009}}.

\bibitem{Lalak:2007vi}
\bibinfo{author}{\bibfnamefont{Z.}~\bibnamefont{Lalak}},
  \bibinfo{author}{\bibfnamefont{D.}~\bibnamefont{Langlois}},
  \bibinfo{author}{\bibfnamefont{S.}~\bibnamefont{Pokorski}}, \bibnamefont{and}
  \bibinfo{author}{\bibfnamefont{K.}~\bibnamefont{Turzynski}},
  \bibinfo{journal}{JCAP} \textbf{\bibinfo{volume}{0707}}, \bibinfo{pages}{014}
  (\bibinfo{year}{2007}), \eprint{0704.0212}.

\bibitem{Liddle:1994dx}
\bibinfo{author}{\bibfnamefont{A.~R.} \bibnamefont{Liddle}},
  \bibinfo{author}{\bibfnamefont{P.}~\bibnamefont{Parsons}}, \bibnamefont{and}
  \bibinfo{author}{\bibfnamefont{J.~D.} \bibnamefont{Barrow}},
  \bibinfo{journal}{Phys. Rev.} \textbf{\bibinfo{volume}{D50}},
  \bibinfo{pages}{7222} (\bibinfo{year}{1994}), \eprint{astro-ph/9408015}.

\bibitem{Mortonson:2014bja}
\bibinfo{author}{\bibfnamefont{M.~J.} \bibnamefont{Mortonson}}
  \bibnamefont{and} \bibinfo{author}{\bibfnamefont{U.}~\bibnamefont{Seljak}}
  (\bibinfo{year}{2014}), \eprint{1405.5857}.

\bibitem{Flauger:2014qra}
\bibinfo{author}{\bibfnamefont{R.}~\bibnamefont{Flauger}},
  \bibinfo{author}{\bibfnamefont{J.~C.} \bibnamefont{Hill}}, \bibnamefont{and}
  \bibinfo{author}{\bibfnamefont{D.~N.} \bibnamefont{Spergel}}
  (\bibinfo{year}{2014}), \eprint{1405.7351}.

\end{thebibliography}
%=============================================================================

\end{document}